\documentstyle[aps,prl,preprint,tighten]{revtex}
\newcommand{\fft}[2]{{#1 \over #2}}
\newcommand{\ft}[2]{{\textstyle{{\scriptstyle #1}\over {\scriptstyle #2}}}}
\newcommand{\half}{{\textstyle{1\over2}}}
\newcommand{\ha}{{1\over2}}

\newcommand{\Tr}{\mbox{Tr\,}}

\newcommand{\beq}{\begin{equation}}
\newcommand{\eeq}[1]{\label{#1}\end{equation}}
\newcommand{\bea}{\begin{eqnarray}}
\newcommand{\eea}[1]{\label{#1}\end{eqnarray}}

\newcommand{\rtitle}[1]{\ifpreprintsty{\sl #1},\fi}

\newcommand{\dalemb}[2]{{\vbox{\hrule height .#2pt
        \hbox{\vrule width.#2pt height#1pt \kern#1pt
                \vrule width.#2pt}
        \hrule height.#2pt}}}
\newcommand{\square}{\mathord{\dalemb{6.8}{7}\hbox{\hskip1pt}}}
\newcommand{\dslash}{{/\kern-6pt\partial}}

\begin{document}
\preprint{\vtop{\hbox{MCTP-02-39}\hbox{hep-th/0207003}\vskip24pt}}

\title{Complementarity of the Maldacena and Karch-Randall Pictures}

\author{M.~J.~Duff$\,{}^{1,*}$, James T.~Liu$\,{}^{1,*}$
and H.~Sati$\,{}^{1,2,}$\footnote{Email addresses:
mduff@umich.edu, jimliu@umich.edu, hsati@maths.adelaide.edu.au}}

\address{${}^1$Michigan Center for Theoretical Physics\\
Randall Laboratory, Department of Physics, University of Michigan\\
Ann Arbor, MI 48109--1120, USA}

\address{\strut\\
${}^2$Department of Pure Mathematics and Department of Physics\\
University of Adelaide\\
Adelaide, SA 5005, Australia}

\maketitle

\begin{abstract}
We perform a one-loop test of the holographic interpretation of the
Karch-Randall model, whereby a massive graviton appears on an AdS$_4$
brane in an AdS$_5$ bulk.  Within the AdS/CFT framework, we examine
the quantum corrections to the graviton propagator on the brane, and
demonstrate that they induce a graviton mass in exact agreement with the
Karch-Randall result.  Interestingly enough, at one loop order, the spin
$0$, spin $1/2$ and spin $1$ loops contribute to the dynamically generated
(mass)$^{2}$ in the same $1: 3: 12$ ratio as enters the Weyl anomaly and
the $1/r^{3}$ corrections to the Newtonian gravitational potential.
\end{abstract}


\narrowtext

%%%%%%%%%%%%%%%%%%%%%%%%%%%%%%%%%%%%%%%%%%%%%%%%%%%%%%%%%%%%%%%%%%%%%%%%%%%%%%
%% Introduction

\section{Introduction}

An old question is whether the graviton could have a small but
non-zero rest mass.  If so, it is unlikely to be described by the
explicit breaking of general covariance that results from the addition
of a Pauli-Fierz mass term to the Einstein Lagrangian.  This gives
rise to the well-known Van Dam-Veltman-Zakharov \cite{vdv,zak}
discontinuity problems in the massless limit, that come about by
jumping from five degrees of freedom to two.  Moreover, recent
attempts \cite{Porrativvz,kogan} to circumvent the discontinuity in the
presence of a non-zero cosmological constant work only at tree level
and the discontinuity re-surfaces%
\footnote{A similar quantum discontinuity arises in the ``partially
massless'' limit as a result of jumping from five degrees of freedom
to four\cite{duff:dls}.}
at one loop \cite{duff:ddls}.  On the other hand, in analogy with
spontaneously broken gauge theories, one might therefore prefer a
dynamical breaking of general covariance, which would be expected
to yield a smooth limit.  However, a conventional Higgs mechanism, in
which a scalar field acquires a non-zero expectation value, does not yield
a mass for the graviton.  The remaining possibility is that the graviton
acquires a mass dynamically and that the would-be Goldstone boson is a
{\it spin one bound state}.  Just such a possibility was suggested in
1975 \cite{Duffdyna}.

Interestingly enough, the idea of a massive graviton arising from a spin
one bound state Goldstone boson has recently been revived by Porrati
\cite{Porrati:2001db} in the context of the Karch-Randall brane-world
\cite{karch} whereby our universe is an AdS$_4$ brane embedded in an
AdS$_5$ bulk.  This model predicts a small but finite
four-dimensional graviton mass
\begin{equation}
M^2=\frac{3L_5^2}{2L_4^4},
\label{eq:gmass}
\end{equation}
in the limit $L_4\to\infty$, where $L_4$ and $L_5$ are the `radii' of
AdS$_4$ and AdS$_5$, respectively.  From the Karch-Randall point of
view, the massive graviton bound to the brane arises from solving the
classical $D=5$ linearized gravity equations in the brane background
\cite{karch}.  Furthermore, holography of the Karch-Randall model
\cite{Porrati:2001gx,Bousso:2001cf} consistently predicts an identical
graviton mass.

In a previous paper \cite{Duffliu}, the complementarity
between the Maldacena AdS/CFT correspondence
\cite{Maldacena,Wittenads,Gubserklebanovpolyakov} and the Randall-Sundrum
\cite{Randall} Minkowski braneworld picture was put to the test by
calculating the $1/r^{3}$ corrections to the Newtonian gravitational
potential arising from the CFT loop corrections to the graviton
propagator.  At one loop we have \cite{Duff2}
\begin{equation}
V(r)= \frac{G_{4}m_{1}m_{2}}{r}\biggl
(1+\frac{\alpha G_{4}}{r^{2}} \biggr),
\end{equation}
where $G_{4}$ is the four-dimensional Newton's constant,
\begin{equation}
\alpha=\frac{1}{45\pi}(12n_{1}+3n_{1/2}+n_{0}),
\label{alpha}
\end{equation}
and where $n_0$, $n_{1/2}$ and $n_1$ count the number of (real) scalars,
(Majorana) spinors and vectors in the multiplet. The coefficient $\alpha$
is the same one that determines that part of the Weyl anomaly
involving the square of the Weyl tensor \cite{Duffweyl}.  The fields
on the brane are given by ${\cal N}=4$ supergravity coupled to a
${\cal N}=4$ super-Yang-Mills CFT with gauge group $U(N)$, for which
$(n_{1},n_{1/2},n_{0})=(N^{2},4N^{2},6N^{2})$.  Using both the AdS/CFT
relation, $N^{2}=\pi L_{5}{}^{3}/2G_{5}$, and the brane world
relation, $G_{4}=2G_5/L_{5}$, we find
\begin{equation}
G_{4}\alpha=\frac{G_{4}L_{5}{}^{3}}{3G_{5}}=\frac{2L_{5}{}^{2}}{3},
\label{result}
\end{equation}
where $G_{5}$ is the five-dimensional Newton's constant.  Hence
\begin{equation}
V(r)= \frac{G_{4}m_{1}m_{2}}{r}\biggl(1+\frac{2L_{5}{}^{2}}{3r^{2}} \biggr),
\label{Newton}
\end{equation}
which agrees exactly with the Randall-Sundrum bulk result.

This complementarity can be generalized to the Karch-Randall AdS braneworld
picture.  From an AdS/CFT point of view, one may equally well foliate
a Poincar\'e patch of AdS$_5$ in AdS$_4$ slices.  The Karch-Randall
brane is then such a slice that cuts off the AdS$_5$ bulk.  However,
unlike for the Minkowski braneworld, this cutoff is not complete, and
part of the original AdS$_5$ boundary remains
\cite{karch,Bousso:2001cf}.  Starting with a maximally supersymmetric
gauged ${\cal N}=8$ supergravity in the five dimensional bulk, the
result is a gauged ${\cal N}=4$ supergravity on the brane coupled to a
${\cal N}=4$ super-Yang-Mills CFT with gauge group $U(N)$, however
with unusual boundary conditions on the CFT fields
\cite{Porrati:2001gx,Bousso:2001cf,DeWolfe:2001pq,Porrati:2001db,Erdmenger:2002ex}.

As was demonstrated in Ref.~\cite{Porrati:2001db}, the CFT on AdS$_4$
provides a natural origin for the bound state Goldstone boson which
turns out to correspond to a {\it massive} representation of $SO(3,2)$.
However, while Ref.~\cite{Porrati:2001db} considers the case of
coupling to a single conformal scalar, in this paper we provide a
crucial test of the complementarity by computing the dynamically
generated graviton mass induced by a complete ${\cal N}=4$
super-Yang-Mills CFT on the brane and showing that this quantum
computation correctly reproduces the Karch-Randall result,
(\ref{eq:gmass}).

We begin in section 2 by discussing properties of the graviton propagator
and providing a general framework for the dynamical generation of graviton
mass.  In section 3, we introduce homogeneous coordinates, and set up
the loop computation, which we carry out in section 4.  Finally, in
section 5, we recover the Karch-Randall graviton mass, (\ref{eq:gmass}),
from the from the quantum CFT perspective.

%%%%%%%%%%%%%%%%%%%%%%%%%%%%%%%%%%%%%%%%%%%%%%%%%%%%%%%%%%%%%%%%%%%%%%%%%%%%%%
%% Transverse self-energy

\section{Transversality and the graviton mass}

We are mainly interested in the properties of the
one-loop graviton self-energy, $\Sigma_{\mu\nu,\alpha\beta}(x,y)$.
As emphasized in Refs.~\cite{Duffdyna,Porrati:2001db}, mass generation is
compatible with the gravitational Ward identity arising from
diffeomorphism invariance.  Thus the self-energy remains transverse,
$\nabla_x^\mu\Sigma_{\mu\nu,\alpha\beta}=
\nabla_y^\alpha\Sigma_{\mu\nu,\alpha\beta}=0$.  One is then
able to write $\Sigma$ as a non-local expression evaluated at point
$x^\mu$, compatible with transversality
\begin{equation}
\Sigma_{\mu\nu,\alpha\beta}(x)=\beta(\Delta)\Pi^{}_{\mu\nu,\alpha\beta}
(\Delta)+\gamma(\Delta)K_{\mu\nu,\alpha\beta}(\Delta) ,
\label{eq:selfe}
\end{equation}
where \cite{Porrati:2001db}
\widetext
\begin{eqnarray}
\Pi_{\mu\nu}^{}{}^{\alpha\beta}&=&\delta_\mu^\alpha\delta_\nu^\beta
-\fft13g_{\mu\nu}g^{\alpha\beta}
+2\nabla_{\mu}\left(\fft{\delta_\nu^{\beta}+\nabla_{\nu}\nabla^{\beta}/2\Lambda}
{\Delta-2\Lambda}\right)\nabla^{\alpha}
\nonumber\\
&&\qquad
-\fft\Lambda3(g_{\mu\nu}+\fft3\Lambda\nabla_{\mu}\nabla_{\nu})\fft1{3\Delta
-4\Lambda}(g^{\alpha\beta}+\fft3\Lambda\nabla^{\alpha}\nabla^{\beta})
\label{eq:pidef}
\end{eqnarray}
\narrowtext
\noindent
is the transverse-traceless projection and
\begin{equation}
K_{\mu\nu}{}^{\alpha\beta}=\fft{\Delta-\Lambda}{3\Delta-4\Lambda}
d_{\mu\nu}d^{\alpha\beta};\quad 
d_{\mu\nu}=g_{\mu\nu}+\fft1{\Delta-\Lambda}\nabla_{\mu}\nabla_{\nu}
\end{equation}
is the transverse but trace projection.  More generally,
\begin{equation}
(\Pi+K)_{\mu\nu}{}^{\alpha\beta}=\delta_\mu^\alpha\delta_\nu^\beta
+\fft2{\Delta-2\Lambda}\delta_\mu^\alpha\nabla_\nu\nabla^\beta
+\fft1{(\Delta-2\Lambda)(\Delta-\Lambda)}\nabla_\mu\nabla_\nu
\nabla^\alpha\nabla^\beta,
\end{equation}
is an overall transverse projection, regardless of trace.
Here, $\Lambda=-3/L_{4}^{2}$ is
the four-dimensional cosmological constant and $\Delta$ is the general
Lichnerowicz operator which commutes with covariant derivatives.  
Symmetrization on $(\mu\nu)$ and $(\alpha\beta)$ is implied throughout.
In flat space, these expressions reduce simply to the familiar
\begin{equation}
\Pi_{\mu\nu}{}^{\alpha\beta}=d_\mu^\alpha d_\nu^\beta-\ft13d_{\mu\nu}
d^{\alpha\beta},\qquad
K_{\mu\nu}^{\alpha\beta}=\ft13d_{\mu\nu}d^{\alpha\beta}
\end{equation}
where
\begin{equation}
d_{\mu\nu}=\eta_{\mu\nu}-\fft{\partial_\mu\partial_\nu}{\square}.
\end{equation}

In Feynman gauge, the tree-level massless graviton propagator in AdS
takes the form
\begin{equation}
D_{\mu\nu}{}^{\alpha\beta}=\fft1{\Delta-2\Lambda}(\delta_\mu^\alpha
\delta_\nu^\beta-\ft12g_{\mu\nu}g^{\alpha\beta}).
\end{equation}
Using the self-energy written in the form (\ref{eq:selfe}), the quantum
corrected propagator may be summed to yield
\begin{eqnarray}
\widetilde D_{\mu\nu}{}^{\alpha\beta}
=\fft1{\Delta-2\Lambda-\beta}\left(\delta_\mu^\alpha\delta_\nu^\beta
-\fft{\Delta-\Lambda}{3\Delta-4\Lambda}
g_{\mu\nu}g^{\alpha\beta}\right)
-\fft1{\Delta-\Lambda+\gamma/2}
\left(\fft12\fft{\Delta-\Lambda}{3\Delta-4\Lambda}
g_{\mu\nu}g^{\alpha\beta}\right)
\label{eq:rsprop}
\end{eqnarray}
when evaluated between conserved sources.  This indicates that a constant
piece in the traceless self-energy, $\beta=-M^2$, will shift the spin-2
pole in the propagator, thus
yielding a non-zero graviton mass.  The second term, involving the 
trace, may combine with the scalar part of the first.  However a 
potentially dangerous scalar ghost pole at $3\Delta=4\Lambda$ may appear.
This ghost is absent whenever the residue of the pole vanishes, {\it i.e.}
provided 
$\left.\gamma=\beta\right|_{4\Delta=3\Lambda}$.  This is in fact the 
case, as may be seen by explicit computation below.  Although the 
field theory is conformal, the presence of $K$ is demanded by the Weyl 
anomaly \cite{Duffweyl}.  However, this trace piece is entirely 
contained in the local part of $\Sigma$, and does not contribute 
directly to the mass. The net result is a pure massive spin-2 
propagator
\begin{eqnarray}
\widetilde D_{\mu\nu}{}^{\alpha\beta}
=\fft1{\Delta-2\Lambda+M^2}\biggl(\delta_\mu^\alpha\delta_\nu^\beta
-\fft12\left(\fft{2\Lambda-2M^2}{2\Lambda-3M^2}\right)
g_{\mu\nu}g^{\alpha\beta}\biggr),
\label{eq:ms2prop}
\end{eqnarray}
where we have taken $\beta=-M^2$ and the Pauli-Fierz combination, 
$\gamma=(3/\Lambda)(\Delta-\Lambda)\beta$.

Thus the procedure we follow in determining the graviton mass is to
compute the one-loop self-energy in an AdS background, and to identify
the appropriate constant piece $\beta$.  Viewed in coordinate space,
this is a non-local contribution to $\Sigma$.  But this is precisely
what is necessary to induce a graviton mass.

%%%%%%%%%%%%%%%%%%%%%%%%%%%%%%%%%%%%%%%%%%%%%%%%%%%%%%%%%%%%%%%%%%%%%%%%%%%%%%
%% Homogeneous coordinates and bi-tensors

\section{Homogeneous coordinates and bi-tensors}

Before turning to an explicit calculation of the graviton self energy,
we consider some preliminaries for studying quantum fields in
homogeneous spaces.  In particular, we establish our notation and
review some useful facts about manipulating tensors in homogeneous
space.  Many of these techniques are by now standard; further details
may be found in, {\it e.g.},
Refs.~\cite{Fronsdal:1978vb,Allen:wd,Allen:1986tt,Turyn:1988af,D'Hoker:1999jc}.

We find it convenient to work in homogeneous coordinates,
which corresponds to the embedding of AdS$_4$ in $R^5$ with pseudo-Euclidean
metric, $\eta_{MN}=\,{\rm diag}(-,+,+,+,-)$.  AdS$_4$ is then given by the
restriction to the hyperboloid $X^MX^N\eta_{MN}=-L_{4}^2$.  Note that we denote
homogeneous coordinates as $X^M, Y^M,\ldots$ ($M,N=0,\ldots,4$) and intrinsic
coordinates as $x^\mu,y^\mu,\ldots$ ($\mu,\nu=0,\ldots,3$).

Tensor fields $\phi_{MNP\cdots}(X)$ restricted to the hyperboloid must
satisfy $X^M\phi_{MNP\cdots}(X)=0$.  In addition, we take them to be
homogeneous of (arbitrary) degree $n$, $\phi_{MNP\cdots}(\lambda X)=
\lambda^n\phi_{MNP\cdots}(X)$.  An important point to note
in transforming from intrinsic coordinates to homogeneous
coordinates is that all tensor indices must be restricted to lie on the
hyperboloid, namely $X^M\phi_{MNP\cdots}(X)=0$.  Projecting into the
tangent direction at a point $X^M$ is accomplished by the operator
\begin{equation}
G_{MN}(X)=\eta_{MN}+X_MX_N/L^2
\label{eq:gee}
\end{equation}
which also serves as a metric tensor where $\Tr(G)\equiv G_{MN}G^{MN}=4$
(recall that $X^2=-L^2$).

Two-point functions in coordinate space are in general bi-tensor
functions of the points $X^M$ and $Y^P$.  Maximally symmetric scalar
functions, $\phi(X,Y)$, are simple and can only depend on the invariant
$|X-Y|^2/L^2=-2(Z+1)$ where $Z=X\cdot Y/L^2$.  However, in general, we
must also consider bi-tensors of the form, $\phi_{MN\cdots,PQ\cdots}(X,Y)$,
where the first (second) set of indices refer to point $X^M$ ($Y^P$).
To construct such bi-tensors, we define the unit vectors
\begin{equation}
N_M(X)=\fft{Y_M+ZX_M}{L\sqrt{Z^2-1}},\qquad
N_P(Y)=\fft{X_P+ZY_P}{L\sqrt{Z^2-1}}
\label{eq:uvec}
\end{equation}
where, as before, $Z=X\cdot Y/L^2$.  These serve the same purpose as the
unit tangent vectors of \cite{Allen:wd}, except that here they are given
in homogeneous coordinates.  In addition, we also make use
of the mixed tensor
\begin{equation}
\hat G_{MP}(X,Y)=G_{MN}(X)\eta^{NQ}G_{PQ}(Y)
=\eta_{MP}+(X_MX_P+Y_MY_P+ZX_MY_P)/L^2.
\end{equation}
This serves the same function as the `parallel propagator' of
Ref.~\cite{Allen:wd}.  However, when converted from intrinsic
coordinates, the parallel propagator has the form
$g_{MP}=\hat G_{MP}-(Z+1)N_MN_P$, which differs at large separations.
We choose to use $\hat G_{MP}$ because it is symmetric under the
anti-podal map $Y\to-Y$ in the covering space of AdS, while $g_{MP}$ is not.

It is clear from the condition of maximal symmetry that all bi-tensors
may be expressed in terms of the metrics $G_{MN}(X)$, $G_{PQ}(Y)$, unit
vectors $N_M(X)$, $N_P(Y)$, and mixed tensor $\hat G_{MP}(X,Y)$.  For
the graviton self energy, we are interested in the two point function,
$\langle T_{MN}(X)T_{PQ}(Y)\rangle$.  Since this is
symmetric under either $M\leftrightarrow N$ or $P\leftrightarrow Q$ or the
simultaneous interchange of $MN\leftrightarrow PQ$ and $X\leftrightarrow
Y$, it may always be decomposed in terms of a set of five basis
bi-tensors, which we take to be \cite{Allen:1986tt}
\begin{eqnarray}
&\displaystyle
{\cal O}_1=G_{MN}G_{PQ},\qquad
{\cal O}_2=N_MN_NN_PN_Q,\qquad
{\cal O}_3=2\hat G_{M}{}^{(P}\hat G_{N}{}^{Q)},&\nonumber\\
&\displaystyle
{\cal O}_4=G_{MN}N_PN_Q+N_MN_NG_{PQ},\qquad
{\cal O}_5=4\hat G_{(M}{}^{(P}N_{N)}N^{Q)}.&
\label{eq:o5}
\end{eqnarray}
To avoid lengthening the notation, we do not include the $X$ or $Y$
dependence explicitly; indices $M$ and $N$ always refer to $X$, and $P$
and $Q$ always refer to $Y$.  With all indices contracted against proper
homogeneous tensors, these operators may be represented simply by
\begin{eqnarray}
&\displaystyle
{\cal O}_1=\eta_{MN}\eta^{PQ},\qquad
{\cal O}_2=Y_MY_NX_PX_Q/L^4(Z^2-1)^2,\qquad
{\cal O}_3=2\delta_{(M}^{(P}\delta_{N)}^{Q)}&\nonumber\\
&\displaystyle
{\cal O}_4=(\eta_{MN}X^PX^Q+Y_MY_N\eta^{PQ})/L^2(Z^2-1),\qquad
{\cal O}_5=4\delta_{(M}^{(P}Y_{N)}X^{Q)}/L^2(Z^2-1).&
\end{eqnarray}
These expressions are sufficient for determining the appropriate linear
combinations of the operators without having to keep track of complete
projections.  The complete operators, (\ref{eq:o5}), may be recovered by
projecting all external indices with (\ref{eq:gee}).

Note that this decomposition follows the notation of Ref.~\cite{Allen:wd}
(with tensor quantities have been converted to homogeneous coordinates),
except that we use the mixed tensor $\hat G_{MP}$ instead of the
parallel propagator $g_{MP}$.  This choice leads to more symmetric
expressions, and highlights the interplay between boundary conditions
and the use of image charges below.  In terms of the parallel
propagator, Ref.~\cite{Allen:1986tt} would define instead
\begin{equation}
\widetilde{\cal O}_3=2 g_{M}{}^{(P}g_{N}{}^{Q)},\qquad
\widetilde{\cal O}_5=4 g_{(M}{}^{(P}N_{N)}N^{Q)},
\end{equation}
instead.  The relation between the two bases is given by
\begin{eqnarray}
{\cal O}_3&=&\widetilde{\cal O}_3+(Z+1)\widetilde{\cal
O}_5+2(Z+1)^2{\cal O}_2,\nonumber\\
{\cal O}_5&=&\widetilde{\cal O}_5+(Z+1){\cal O}_2
\end{eqnarray}
(with the remaining unchanged).  This is a straightforward
identification at short distances ($Z\to-1$), and only differs at long
distances.

In order to investigate the graviton self energy, it is useful to obtain
a basis of transverse traceless bi-tensors.  Although, in the flat space
limit, transversality is easily expressed in momentum space, this is not
the case when working in homogeneous coordinates.  We first define the
three traceless combinations
\begin{eqnarray}
T_1&=&\fft1{3(3Z^2+1)}[{\cal O}_1+16{\cal O}_2-4{\cal O}_4], \nonumber\\
T_2&=&-\ft13{\cal O}_1 +\ft23{\cal O}_2
+\ft12\widetilde{\cal O}_3+\ft13{\cal O}_4+\widetilde\ft12{\cal O}_5,
\nonumber\\
T_3&=&\fft1{2Z}[4{\cal O}_2+\widetilde{\cal O}_5],
\label{eq:ttens}
\end{eqnarray}
where $T_1$, $T_2$, and $T_3$ are traceless in the sense
$G^{MN}T_{MN,PQ}=T_{MN,PQ}G^{PQ}=0$.  For completeness, there is also a
pure trace combination
\begin{equation}
P_R=\fft1{Z^2(3Z^2+1)}[Z^4{\cal O}_1+(Z^2-1)^2{\cal O}_2-Z^2(Z^2-1){\cal
O}_4].
\label{eq:ptens}
\end{equation}
While there is some arbitrariness in the definition of $T_1$, $T_2$ and
$T_3$, the above definitions (including normalization) were chosen to
have a natural reduction in the flat space (or short distance) limit.

This limit corresponds to taking $Y\to X$, so that $Z\to 1$, and both
$G$ and $\hat G$ reduce to the (four-dimensional) flat space metric
$\eta$.  In addition, the tangent vectors, (\ref{eq:uvec}), reduce
according to
\begin{equation}
N_M\to \hat r_\mu,\qquad N_P\to -\hat r_\rho
\end{equation}
where $\hat r = (\vec y-\vec x\,)/|\vec y-\vec x\,|$.  The resulting
traceless (\ref{eq:ttens}) and trace (\ref{eq:ptens}) combinations take
on the projection form
\begin{eqnarray}
T_1&\to&\ft1{12}(\eta_{\mu\nu}-4\hat r_\mu\hat r_\nu)
(\eta^{\rho\sigma}-4\hat r^\rho\hat r^\sigma),\nonumber\\
T_2&\to& (\delta_{(\mu}^{(\rho}-\hat r_\mu\hat r^\rho)
(\delta_{\nu}^{\sigma}-\hat r_{\nu)}\hat r^{\sigma)})
-\ft13(\eta_{\mu\nu}-\hat r_\mu\hat r_\nu)
(\eta^{\rho\sigma}-\hat r^\rho\hat r^\sigma),\nonumber\\
T_3&\to& (\delta_{(\mu}^{(\rho}-\hat r_\mu\hat r^\rho)\hat r_{\nu)}\hat
r^{\sigma)},\nonumber\\
P_R&\to&\ft14\eta_{\mu\nu}\eta^{\rho\sigma}
\end{eqnarray}
These projections are essentially onto longitudinal, transverse
traceless, transverse and pure trace components, with rank
1, 5, 3 and 1, respectively.

Returning to AdS, it should be noted that, while traceless, $T_1$, $T_2$
and $T_3$ are not in themselves transverse.  However, any transverse
traceless bi-tensor must be able to be written as a combination
\begin{equation}
{\cal T}=a_1(Z)(3Z^2+1)T_1+a_2(Z)T_2+a_3(Z)T_3
\end{equation}
where transversality imposes two conditions on the three functions
$a_1$, $a_2$ and $a_3$.  The details are carried out in Appendix B; the
result is that to highlight the large separation behavior of ${\cal T}$,
we construct a basic of transverse traceless bi-tensors $\{{\cal
T}_{(n)}\}$.  Below, when examining the graviton self-energy, we will make
use of this basis for extracting the non-local quantity responsible for
graviton mass generation.

%%%%%%%%%%%%%%%%%%%%%%%%%%%%%%%%%%%%%%%%%%%%%%%%%%%%%%%%%%%%%%%%%%%%%%%%%%%%%%
%% Computation of the self energy

\section{Computation of the graviton self energy}

Before addressing the one-loop computation, we start by examining the
scalar, fermion and vector two-point functions, paying attention
to necessary boundary conditions \cite{Avis:1977yn,Porrati:2001db}.
Details are provided in the Appendix; here we simply summarize the
results.
A normalized scalar propagator necessarily has short-distance behavior
\begin{equation}
\Delta_0(X,Y)\sim -\fft1{4\pi^2}\fft1{|X-Y|^2}\sim
\fft1{8\pi^2L_{4}^2}\fft1{Z+1},
\label{eq:tscal}
\end{equation}
so that it reduces properly in the flat space limit.  However, boundary 
conditions must still be satisfied by the addition of an appropriate
solution to the homogeneous equation.
For AdS energy $E_{0}=1$ or $2$, and for mixed boundary conditions
encoded by parameters $\alpha_+$,
$\alpha_-$, the scalar propagator takes the form \cite{Avis:1977yn}
\begin{equation}
\Delta_0^{(\alpha)}
=\fft1{8\pi^2L_{4}^2}\left(\fft{\alpha_+}{Z+1}+\fft{\alpha_-}{Z-1}\right).
\label{eq:gscal}
\end{equation}
Although normalization demands
$\alpha_+=1$, we nevertheless find it illuminating to keep $\alpha_+$
arbitrary, as it highlights the symmetries in the latter expressions for
the graviton self energy computation.  Note that $\alpha_-=0$ 
corresponds to transparent boundary conditions, while $\alpha=\pm1$ 
corresponds to ordinary reflecting ones.

Similarly, the appropriate fermion propagator has the form
\begin{equation}
\Delta_{1/2}^{(\alpha)}=\fft1{8\pi^2L^4}\left(\alpha_+\fft{\Gamma^M(X_M-Y_M)}
{(Z+1)^2}
+\alpha_-\fft{\Gamma^M(X_M+Y_M)}{(Z-1)^2}\right).
\label{eq:12prop}
\end{equation}
The vector propagator is the first case where we have to worry about
bi-tensor structures as well as gauge fixing.  However, for correlators
of the stress tensor, we only need the expression for the gauge invariant
two-point function $\langle F_{MN}(X)F_{PQ}(Y)\rangle$.  The result is
\begin{eqnarray}
\langle F_{MN}(X)F^{PQ}(Y)\rangle^{(\alpha)}&=&\fft1{2\pi^2L^4}\Biggl[
\fft{\alpha_+}{(Z+1)^2}
(\hat G_{[M}{}^{[P}\hat G_{N]}{}^{Q]}-2(Z-1)N_{[M}\hat G_{N]}{}^{[Q}N^{P]})
\nonumber\\
&&\qquad\quad
+\fft{\alpha_-}{(Z-1)^2}
(\hat G_{[M}{}^{[P}\hat G_{N]}{}^{Q]}-2(Z+1)N_{[M}\hat G_{N]}{}^{[Q}N^{P]})
\Biggr]
\label{eq:1prop}
\end{eqnarray}

\subsection{The scalar contribution}

The scalar loop contribution to the graviton self energy was partially
computed in Ref.~\cite{Porrati:2001db}, where the proper r\^ole of
boundary conditions was highlighted.  The relevant Lagrangian for a
conformally coupled scalar is given by
\begin{equation}
e^{-1}{\cal L}=-\half\partial\phi^2-\ft1{12}R\phi^2.
\end{equation}
This gives rise to the equation of motion, $(\square-\fft16R)\phi=0$, as
well as the improved stress tensor
\begin{equation}
T_{\mu\nu}=\partial_\mu\phi\partial_\nu\phi-\half g_{\mu\nu}(\partial\phi)^2
-\ft16[\nabla_\mu\nabla_\nu-g_{\mu\nu}\square-(R_{\mu\nu}-\half
g_{\mu\nu}R)]\phi^2.
\label{eq:stmunu}
\end{equation}
This stress tensor is both conserved and traceless (on the equations of
motion), as expected for a conformal scalar.  For computational
purposes, we find it convenient to reexpress (\ref{eq:stmunu}) as
\begin{equation}
T_{\mu\nu}=\ft23\partial_\mu\phi\partial_\nu\phi-\ft16g_{\mu\nu}
[(\partial\phi)^2+\phi^2/L^2]-\ft13\phi\nabla_\mu\nabla_\nu\phi
\end{equation}
where we have fixed the background AdS metric and made use of the
scalar equation of motion%
\footnote{Since the induced graviton mass is a long-distance effect, we
are unconcerned with any contact terms that may be discarded by
evaluation of the equation of motion on the Green's functions.  In any
case, such issues may be avoided by, {\it e.g.}, use of a point splitting
regulator.}.

Evaluation of $\langle T_{\mu\nu}(x)T_{\rho\sigma}(y)\rangle$ follows
from Wick's theorem
\begin{eqnarray}
\langle T_{\mu\nu}(x)T_{\rho\sigma}(y)\rangle
&=&\ft49\langle\partial_\mu\phi(x)\partial_\nu\phi(x)\partial_\rho\phi(y)
\partial_\sigma\phi(y)\rangle+\cdots\nonumber\\
&=&\ft49[\partial_\mu\partial_\rho\Delta_0(x-y)\partial_\nu\partial_\sigma
\Delta_0(x-y)+\partial_\mu\partial_\sigma\Delta_0(x-y)\partial_\nu
\partial_\rho\Delta_0(x-y)]+\cdots,\nonumber\\
\end{eqnarray}
where $\Delta_0$ is the scalar propagator.
Working in homogeneous coordinates, after considerable manipulation, we
obtain the self energy as a bi-local tensor
\begin{eqnarray}
\langle T_{MN}(X)T_{PQ}(Y)\rangle
&=&\hphantom{+}
{\cal O}_1[\ft1{18}(Z\Delta_0'+\Delta_0)^2-\ft{11}{18}\Delta_0'^2
+\ft19\Delta_0\Delta_0'']\nonumber\\
&&+{\cal O}_2(Z^2-1)^2[\Delta_0''^2+\ft19\Delta_0\Delta_0''''
-\ft89\Delta_0'\Delta_0''']\nonumber\\
&&+{\cal O}_3[\ft49\Delta_0'^2+\ft19\Delta_0\Delta_0'']\nonumber\\
&&+{\cal O}_4(Z^2-1)[-\ft{14}9\Delta_0'^2+\ft79\Delta_0\Delta_0''
+\ft19Z\Delta_0\Delta_0'''-\ft13Z\Delta_0'\Delta_0'']\nonumber\\
&&+{\cal O}_5(Z^2-1)[\ft19\Delta_0\Delta_0'''],
\end{eqnarray}
where primes denote differentiation with respect to $Z$.  Note that, to
simplify the expression, we have used the scalar equation of motion,
$(Z^2-1)\Delta_0''=-2(\Delta_0+2Z\Delta_0')$, where we have dropped the
short-distance term $\delta(X-Y)$.  Substituting in the explicit form of
the scalar propagator, (\ref{eq:gscal}), we find
\begin{eqnarray}
\langle T_{MN}(X)T_{PQ}(Y)\rangle_0&=&\fft1{48\pi^4L_{4}^8}\Biggl[
\fft{\alpha_+^2}{(Z+1)^4} \left(\fft{3Z^2+1}4T_1+T_2+ZT_3\right) \nonumber\\
&&\qquad+\fft{\alpha_-^2}{(Z-1)^4} \left(\fft{3Z^2+1}4T_1+T_2-ZT_3\right)
\nonumber\\
&&\qquad+\fft23\fft{\alpha_+\alpha_-}{(Z^2-1)^3}
\bigl(5(3Z^2+1)T_1+(3Z^2-1)T_2-10Z^2T_3\bigr)\Biggr]
\label{eq:0se}
\end{eqnarray}
(up to contact terms, which we drop).

\subsection{The fermion contribution}

Turning next to spin-1/2, we take for simplicity a massless Dirac
fermion with Lagrangian
\begin{equation}
e^{-1}{\cal L}=\half\overline{\psi}(\stackrel{\rightarrow}{/\kern-6pt\nabla}
-\stackrel{\leftarrow}{/\kern-6pt\nabla})\psi
\end{equation}
and stress tensor
\begin{equation}
T_{\mu\nu}=\half\overline{\psi}\gamma_{(\mu}(\stackrel{\rightarrow}{\nabla_\nu}
-\stackrel{\leftarrow}{\nabla_{\nu)}})\psi-\half g_{\mu\nu}
\overline{\psi}(\stackrel{\rightarrow}{/\kern-6pt\nabla}
-\stackrel{\leftarrow}{/\kern-6pt\nabla})\psi
\label{eq:12tmn}
\end{equation}
Note that this is both traceless and covariantly conserved on the
equations of motion, $/\kern-6pt\nabla\psi=\overline{\psi}
\stackrel{\leftarrow}{/\kern-6pt\nabla}=0$.

As in the scalar case, use of the equations of motion on the external
vertices allows us to ignore the second term in (\ref{eq:12tmn}) when
evaluating $\langle T_{\mu\nu}(x)T_{\rho\sigma}(y)\rangle$.  Promoting
this expression to homogeneous coordinates, and using Wick's theorem, we
find
\begin{equation}
\langle T_{MN}(X)T_{PQ}(Y)\rangle=-\ft14\Tr[\Gamma_{(M}(\stackrel{\rightarrow}
{\partial_N}-\stackrel{\leftarrow}{\partial_{N)}})\Delta_{1/2}(X,Y)
\Gamma_{(P}(\stackrel{\rightarrow}
{\partial_Q}-\stackrel{\leftarrow}{\partial_{Q)}})\Delta_{1/2}(Y,X)]
\end{equation}
(the $-$ sign is for a fermion loop)
where $\Delta_{1/2}(X,Y)$ is the spin-1/2 propagator given in
(\ref{eq:12prop}).  The Dirac trace may be evaluated by writing
$\Delta_{1/2}(X,Y)=\Gamma_A\Delta_{1/2}^A(X,Y)$, so that
\begin{eqnarray}
\langle T_{MN}(X)T_{PQ}(Y)\rangle&=&-(\delta_M^A\delta_P^B+\delta_P^A\delta_M^B
-\eta_{MP}\eta^{AB})\nonumber\\
&&\times\bigl[\partial_N\Delta_{1/2}^A(X,Y)\partial_Q\Delta_{1/2}^B(Y,X)
+\partial_Q\Delta_{1/2}^A(X,Y)\partial_N\Delta_{1/2}^B(Y,X)\nonumber\\
&&\quad -\Delta_{1/2}^A(X,Y)\partial_N\partial_Q\Delta_{1/2}^B(Y,X)
-(\partial_N\partial_Q\Delta_{1/2}^A(X,Y))\Delta_{1/2}^B(Y,X)\bigr]
\end{eqnarray}
where a further symmetrization on $(MN)$ and $(PQ)$ is implied.  This
expression is symmetric under interchange of $X\leftrightarrow Y$ in the
propagators.  Since this corresponds to taking
$\alpha_+\leftrightarrow-\alpha_+$ [as is evident from
(\ref{eq:12prop})], the overall result is to project onto terms even in
$\alpha_+$.  In particular, this kills any possible terms proportional
to $\alpha_+\alpha_-$ in the two-point function.

A straightforward computation results in the expression
\begin{eqnarray}
\langle T_{MN}(X)T_{PQ}(Y)\rangle&=&-\fft1{32\pi^4L^8}\Biggl[
\fft{\alpha_+^2}{(Z+1)^4}
({\cal O}_1-2{\cal O}_3+2(Z-1){\cal O}_5
-4(Z-1)^2{\cal O}_2)\nonumber\\
&&\qquad\qquad+\fft{\alpha_-^2}{(Z-1)^4}
({\cal O}_1-2{\cal O}_3+2(Z+1){\cal O}_5
-4(Z+1)^2{\cal O}_2)\Biggr]
\label{eq:fcon}
\end{eqnarray}
which may be rewritten in terms of the traceless $T$ tensors of
(\ref{eq:ttens}) as
\begin{eqnarray}
\langle T_{MN}(X)T_{PQ}(Y)\rangle_{1/2}&=&\fft1{8\pi^4L^8}\Biggl[
\fft{\alpha_+^2}{(Z+1)^4}\left(\fft{3Z^2+1}4T_1+T_2+ZT_3\right)
\nonumber\\
&&\qquad+\fft{\alpha_-^2}{(Z-1)^4}\left(\fft{3Z^2+1}4T_1+T_2-ZT_3\right)
\Biggr].
\end{eqnarray}
Other than for the absence of the mixed $\alpha_-\alpha_+$ term, this
contribution for a Dirac fermion is identical to that of a scalar loop,
(\ref{eq:0se}), but six times larger.  For a Majorana fermion, this
should be halved, so that the contribution is three times that of a
scalar.

\subsection{The vector contribution}

The remaining contribution to the graviton self energy arises from
vector loops.  For a massless gauge boson with Lagrangian
\begin{equation}
e^{-1}{\cal L}=-\ft14F_{\mu\nu}^2,
\end{equation}
the stress tensor is simply
\begin{equation}
T_{\mu\nu}=F_{\mu\lambda}F_\nu{}^\lambda-\ft14g_{\mu\nu}F^2.
\end{equation}
Converting all expressions to homogeneous coordinates, we need to
evaluate
\begin{eqnarray}
\langle T_{MN}(X)T_{PQ}(Y)\rangle&=&2\langle F_M{}^AF_P{}^B\rangle
\langle F_N{}^AF_Q{}^B\rangle-\ft12\eta_{MN}\langle F_C{}^AF_P{}^B\rangle
\langle F^{CA}F_Q{}^B\rangle\nonumber\\
&&-\ft12\eta_{PQ}\langle F_M{}^AF_C{}^B\rangle\langle F_N{}^AF^{CB}\rangle
+\ft18\eta_{MN}\eta_{PQ}\langle F_C{}^AF_D{}^B\rangle\langle
F^{CA}F^{DB}\rangle,
\label{eq:ttff}
\end{eqnarray}
where again symmetrization in $(MN)$ and $(PQ)$ is assumed.  Note that
all contractions are performed with either $G_{MN}(X)$ or $G_{PQ}(Y)$.
The main difficulty is in evaluating the first term in this expression;
the remaining ones follow simply from tracing over the appropriate
indices.  Using the explicit form for the vector propagator,
(\ref{eq:1prop}), we obtain
\begin{eqnarray}
\langle F_M{}^AF_P{}^B\rangle\langle F_N{}^AF_Q{}^B\rangle &=&
\fft1{16\pi^4L^8}\Biggl[\fft{\alpha_+^2}{(Z+1)^4}({\cal O}_1+{\cal
O}_3-(Z-1){\cal O}_5+2(Z-1)^2{\cal O}_2)\nonumber\\
&&\qquad\qquad+\fft{\alpha_-^2}{(Z-1)^4}({\cal O}_1+{\cal
O}_3-(Z+1){\cal O}_5+2(Z+1)^2{\cal O}_2)\nonumber\\
&&\qquad\qquad+\fft{\alpha_+\alpha_-}{(Z^2-1)^2}({\cal O}_1-2{\cal
O}_4)\Biggr].
\end{eqnarray}
Substituting this into (\ref{eq:ttff}), we find that the mixed
$\alpha_+\alpha_-$ term vanishes.  The result is identical to the
fermion case, (\ref{eq:fcon}), except that it is twice as large (as for
the Dirac fermion).  Explicitly, this is given by
\begin{eqnarray}
\langle T_{MN}(X)T_{PQ}(Y)\rangle_1&=&\fft1{4\pi^4L^8}\Biggl[
\fft{\alpha_+^2}{(Z+1)^4}\left(\fft{3Z^2+1}4T_1+T_2+ZT_3\right)\nonumber\\
&&\qquad+\fft{\alpha_-^2}{(Z-1)^4}\left(\fft{3Z^2+1}4T_1+T_2-ZT_3\right)
\Biggr]
\end{eqnarray}

\subsection{The complete supermultiplet}

Until now, we have treated spins 0, $\fft12$ and 1 separately.  However, to
preserve supersymmetry, the boundary conditions on all fields in the
multiplet have to be chosen consistently \cite{Breitenlohner:jf}.  This
means a single set of $\alpha_+$ (actually always 1) and $\alpha_-$
suffices for specifying the boundary conditions.  Furthermore, for a
complex scalar in a Wess-Zumino multiplet, the scalar and
pseudoscalar transform with opposite boundary conditions (even when
the parity condition is relaxed).  Since this corresponds to opposite signs
for $\alpha_-$ between the scalar and pseudoscalar, we see that the
mixed term in (\ref{eq:0se}) always drops out when considering pairs of
spin-0 states as members of supermultiplets.  As a result, we find the
simple universal structure for the graviton self-energy
\begin{eqnarray}
\Sigma_{MN,PQ}(X,Y)&=&8\pi G_4\langle T_{MN}(X)T_{PQ}(Y)\rangle\nonumber\\
&=&8\pi G_4\fft{n_0+3n_{1/2}+12n_1}{48\pi^4L_{4}^8}
\Biggl[\fft{\alpha_+^2}{(Z+1)^4} \left(\fft{3Z^2+1}4T_1+T_2+ZT_3\right)
\nonumber\\
&&\qquad
+\fft{\alpha_-^2}{(Z-1)^4} \left(\fft{3Z^2+1}4T_1+T_2-ZT_3\right) \Biggr].
\label{eq:seans}
\end{eqnarray}
\narrowtext
\noindent
%

%%%%%%%%%%%%%%%%%%%%%%%%%%%%%%%%%%%%%%%%%%%%%%%%%%%%%%%%%%%%%%%%%%%%%%%%%%%%%%
%% Determination of the mass

\section{Extraction of the graviton mass}

We now extract the induced graviton mass from the long distance
behavior of the self energy (\ref{eq:seans}).  We first note that the 
three terms of $\Pi$ in Eq.~(\ref{eq:pidef}) correspond to local 
tensor, non-local spin-1 and spin-0 exchange, respectively.  Following
Ref.~\cite{Porrati:2001db}, the mass can be read off by identifying
in $\Sigma$ a piece proportional to the spin-1 Goldstone boson exchange,
given by the second term in Eq.~(\ref{eq:pidef}):
\begin{equation}
\Pi_{\mu\nu\,\alpha\beta}^{\rm (spin\hbox{-}1)} = 
2\nabla_\nu\left(\fft{g_{\nu\beta}+\nabla_\nu\nabla_\beta/2\Lambda}
{\Delta-2\Lambda}\right)\nabla_\alpha = 2\nabla_\mu
D_{\nu\beta}\nabla_\alpha.
\end{equation}
Here, $D_{\mu\nu}$ is the spin-1, $E_0=4$, propagator.

Working in coordinate space, we now rewrite $\Pi^{\rm (spin\hbox{-}1)}$
as a bi-local tensor.  To accomplish this, we start with the homogeneous
space $E_0=4$ vector propagator, which was worked out in Ref.~\cite{Allen:wd}
\begin{eqnarray}
D_{MP}&=&\fft1{48\pi^2L^2}\left[\fft{2(8-15Z^2+9Z^4)}{(Z^2-1)^2}-9Z\log
\fft{Z+1}{Z-1}\right]\hat G_{MP}\nonumber\\
&&+\fft1{48\pi^2L^2}\left[\fft{2Z(-23+24Z^2-9Z^4)}{(Z^2-1)^2}+9(Z^2-1)\log
\fft{Z+1}{Z-1}\right]N_MN_P
\end{eqnarray}
We now freely integrate by parts to obtain
\begin{eqnarray}
\Pi_{MN\,PQ}^{\rm (spin\hbox{-}1)}&=&2\nabla_{X^M}D_{NQ}\nabla_{Y^P}=
-2(\nabla_{X^M}\nabla_{Y^P}D_{NQ})\nonumber\\
&=&
-\fft{2Z}{3\pi^2L_{4}^4(Z^2-1)^3}[5(3Z^2+1)T_1+2T_2-5(Z^2+1)T_3].
\label{eq:pi1prop}
\end{eqnarray}
Note that this non-local $\Pi^{\rm (spin-1)}$ is transverse
and traceless in itself, while the original expression, (\ref{eq:pidef}),
requires an interplay among all terms to ensure transversality.  This
discrepancy arises only through local terms that we have ignored
throughout.

While the one-loop self-energies we have computed all satisfy the
homogeneous coordinate transversality condition, (\ref{eq:ttcond}),
this condition still allows an undetermined $Z$-dependent form factor.
To read off the correctly induced graviton mass, we essentially need to
obtain the constant piece of $\beta(\Delta)$ in (\ref{eq:selfe}),
which may be determined by matching the large $Z$ behavior of
(\ref{eq:seans}) with that of the spin-1 part of $\Pi$, given by
(\ref{eq:pi1prop}).  To do so, we expand both expressions for large $Z$ and
match the asymptotic behavior.  For the self energy, we find
\begin{eqnarray}
\Sigma &=&8\pi G_4 \fft{n_0+3n_{1/2}+12n_1}{48\pi^4L^8}[(\alpha_+^2+\alpha_-^2)
(\ft14{\cal T}_{(4)}+\ft52{\cal T}_{(6)}+\ft{35}4{\cal T}_{(8)}+\cdots)
\nonumber\\
&&\kern8em+(\alpha_+^2-\alpha_-^2)({\cal T}_{(5)}+5{\cal
T}_{(7)}+\cdots)],
\label{eq:sigt}
\end{eqnarray}
while
\begin{equation}
\Pi^{spin\hbox{-}1}=\fft{10}{3\pi^2L^4}[{\cal T}_{(5)}+3{\cal
T}_{(7)}+\cdots],
\label{eq:pit}
\end{equation}
where the basis forms, ${\cal T}_{(n)}$, are given in Appendix B.
Matching the leading ${\cal T}_{(5)}$ term  gives%
\begin{equation}
M^2=8\pi G_4\fft{n_0+3n_{1/2}+12n_1}{160\pi^2L^4}(\alpha_+^2-\alpha_-^2)
\label{eq:mainr}
\end{equation}
This expression is our main result, and generalizes that obtained in
Ref.~\cite{Porrati:2001db}.  Note, however, that this result differs by
a factor of $160$ from that of Ref.~\cite{Porrati:2001db}.  We believe
that this discrepancy arises from three sources.  Firstly, normalization
of the $E_0=4$ scalar propagator is determined by demanding the proper
strength of the short-distance singularity in the flat space limit
\cite{Allen:wd}.  This yields
\begin{eqnarray}
\Delta_0(E_0=4)&=&-\fft1{4\pi^2L^2}\left[\fft{3Z^2-2}{Z^2-1}
-\fft32Z\log\fft{Z+1}{Z-1}\right]\nonumber\\
&\to&\fft1{8\pi^2L^2}\fft1{Z+1} \qquad \hbox{as}\quad Z\to-1
\end{eqnarray}
[compare with Eq.~(\ref{eq:tscal})].  Taking the large separation
limit, $Z\to\infty$, then gives
\begin{equation}
\Delta_0(E_0=4)\sim-\fft1{10\pi^2L^2}\fft1{Z^4}\qquad\hbox{as}\quad
Z\to\infty
\end{equation}
which accounts for a factor of four.  Secondly, without examining the
tensor structure in detail, there is an ambiguity in attributing the
long range structure of the self energy $\Sigma$ to the propagation of a
spin-1 Goldstone boson in $\Pi$.  In particular, both ${\cal
T}_{(5)}$ and ${\cal T}_{(7)}$ of (\ref{eq:sigt}) and (\ref{eq:pit})
have the requisite long range falloff upon integration by parts
\begin{eqnarray}
h\cdot{\cal T}_{(5)}\cdot h&\sim&\fft3{10}\partial_Mh_{MN}\fft{\eta_{NQ}}
{Z^4}\partial_Ph_{PQ}\nonumber\\
h\cdot{\cal T}_{(7)}\cdot h&\sim&-\fft3{25}\partial_Mh_{MN}\fft{\eta_{NQ}}
{Z^4}\partial_Ph_{PQ}
\end{eqnarray}
As a result, both terms would contribute to the coefficient of the
$1/Z^4$ piece, while only the actual combination ${\cal T}_{(5)}+3{\cal
T}_{(7)}$ of Eq.~(\ref{eq:pit}) may be attributed to the induced
graviton mass.  In other words, it is important to match only the leading
${\cal T}_{(5)}$ behavior between $\Sigma$ and $\Pi$ of
Eqs.~(\ref{eq:sigt}) and (\ref{eq:pit}).  Of course, this would have been
immaterial if the asymptotic expansions had been identical.  However in
this case they are not, and this accounts for another factor of five
between our expression and that of Ref.~\cite{Porrati:2001db}.  Finally,
the remaining factor of eight comes in when determining the mass via the
shift in the pole of the resummed propagator, (\ref{eq:rsprop}).  We find
the mass squared to be simply the constant multiplying
$\Pi^{spin\hbox{-}1}$ (up to
a sign).  Since a canonically normalized graviton couples to the stress
tensor with strength $\kappa h^{\mu\nu}T_{\mu\nu}$, and since we do not
include symmetry factors in our coordinate space Feynman rules, we have
simply
\begin{equation}
\Sigma_{MNPQ}(X,Y)=\kappa^2\langle T_{MN}(X)T_{PQ}(Y)\rangle
=8\pi G_4\langle T_{MN}(X)T_{PQ}(Y)\rangle
\end{equation}
We believe this provides a proper accounting for Newton's constant in
the self energy.  Comparing with Ref.~\cite{Porrati:2001db}, this
appears to be the origin of the remaining factor discrepancy.

Note that the spin-0 term in $\Pi$ has a different structure.  
However this term is canceled by the non-local part of $K$.  The 
absence of spin-0 exchange in $\Sigma$ is in agreement with the AdS 
Higgs mechanism \cite{Porrati:2001db}, and yields the massive 
spin-2 propagator (\ref{eq:ms2prop}) without ghosts.

While we have focused on the dynamical breaking of general covariance,
as evidenced by a mass for the graviton, in a supersymmetric
Karch-Randall model, a dynamical breaking of local supersymmetry and
local gauge invariance also occurs, as evidenced by a mass for the
gravitinos and the gauge bosons.

%%%%%%%%%%%%%%%%%%%%%%%%%%%%%%%%%%%%%%%%%%%%%%%%%%%%%%%%%%%%%%%%%%%%%%%%%%%%%%
%% Discussion and conclusions

For the Karch-Randall braneworld \cite{karch}, where the CFT fields are
that of ${\cal N}=4$ $U(N)$ super-Yang-Mills, we substitute
transparent boundary conditions ($\alpha_+=1$, $\alpha_-=0$) into the
expression for the graviton mass, (\ref{eq:mainr}), and find simply
\begin{equation}
M^2=\fft{9G_4}{4L_{4}^4}\alpha,
\label{eq:masse}
\end{equation}
which reproduces exactly the Karch-Randall result of Eq.~(\ref{eq:gmass}) on
using Eq.~(\ref{result}).  Although we focused on the ${\cal N}=4$ SCFT to
relate the coefficient $\alpha$ to the central charge, the result
(\ref{result}) is universal, being independent of which particular CFT
appears in the AdS/CFT correspondence.  This suggests that $\alpha$
plays a universal r\^ole in both the Minkowski and AdS braneworlds, as
indicated in (\ref{eq:masse}) and (\ref{Newton}), and that our result
is robust at strong coupling.  This presumably explains why our
one-loop computation gives the exact Karch-Randall result.  However,
we do not know for certain whether this persists beyond one loop.

\section*{acknowledgments}

We wish to thank M.~Porrati for enlightening discussions.  JTL wishes to
thank D.~Gross, G.~Horowitz, E.~Mottola and J.~Polchinski for discussions,
and acknowledges the hospitality of the KITP and the UCSB Physics
Department where part of this work was done.
This research was supported in part by DOE Grant DE-FG02-95ER40899.

%%%%%%%%%%%%%%%%%%%%%%%%%%%%%%%%%%%%%%%%%%%%%%%%%%%%%%%%%%%%%%%%%%%%%%%%%%%%%%
%% Appendix: Propagators and self energies

\appendix

\section{Propagators in AdS}

Here we collect some information on spin-0, 1/2 and 1 propagators in
homogeneous coordinates.  First recall that the Casimir of $SO(2,3)$
is $Q=\ha L_{MN}^2=E_0(E_0-3)+s(s+1)$ where $E_0$ and $s$ label the
representation $D(E_0,s)$.  Acting on scalars $\phi(X)$, the
operator $Q$ (corresponding to the Casimir) has the form
\begin{equation}
Q=\half L_{MN}^2=-\half(X_M\partial_N-X_N\partial_M)^2
=\hat N(\hat N+3)-X^2\partial^2
\end{equation}
where $\hat N=X\cdot\partial$.  As a result, the scalar Klein-Gordon
equation is simply
\begin{equation}
[\hat N(\hat N+3)-X^2\partial^2-E_0(E_0-3)]\phi(X)=0
\label{eq:kge}
\end{equation}
To obtain the scalar Green's function between points $X$ and $Y$, we
note that $\partial^2=-\partial_Z^2/L^2$ and $\hat N=X\cdot\partial
=Z\partial_Z$.  In this case, we find that $\Delta_0(Z)\equiv\Delta_0(X,Y)$
satisfies the equation
\begin{equation}
[(1-Z^2)\partial_Z^2-4Z\partial_Z+E_0(E_0-3)]\Delta_0(Z)=0
\label{eq:slap}
\end{equation}
For either $E_0=1$ or $2$, this has a simple pair of solutions,
$\Delta_0\sim 1/(Z\pm1)$.  However, in order to reproduce a short distance
behavior $\Delta_0\sim 1/|X-Y|^2$, we must take the one with the positive
sign.  As a result, we obtain
\begin{equation}
\Delta_0=\fft1{8\pi^2L^2}\fft1{Z+1}=-\fft1{4\pi^2}\fft1{|X-Y|^2}
\label{eq:aatscal}
\end{equation}
The normalization is fixed by demanding that $\Delta_0$ reduces properly
in the flat space limit.

The propagator of (\ref{eq:aatscal}) in fact corresponds to imposing
transparent boundary conditions on the scalar.  This is seen by
recalling that while (the covering space of) AdS$_4$ may be conformally
mapped into half of the Einstein static universe, with topology
$R\times S_3$, (\ref{eq:aatscal}) is in fact well defined on the
complete $S_3$ (so that the boundary is in effect invisible)
\cite{Avis:1977yn}%
Reflective boundary conditions may be imposed by
a method of images so that
\begin{equation}
\Delta_0^\pm=-\fft1{4\pi^2}\left(\fft1{|X-Y|^2}\pm\fft1{|X+Y|^2}\right)
=\fft1{8\pi^2L^2}\left(\fft1{Z+1}\mp\fft1{Z-1}\right)
\end{equation}
It is now evident that mixed boundary conditions may be encoded by
parameters $\alpha_+$, $\alpha_-$ where
\begin{equation}
\Delta_0^{(\alpha)}
=\fft1{8\pi^2L^2}\left(\fft{\alpha_+}{Z+1}+\fft{\alpha_-}{Z-1}\right)
\label{eq:aagscal}
\end{equation}
While the residue of the short distance pole must be fixed
({\it i.e.}\ $\alpha_+=1$), we find it illuminating to keep $\alpha_+$
arbitrary, as it highlights the symmetries in the latter expressions for
the graviton self energy computation.  In terms of Porrati's $\alpha$
and $\beta$ coefficients, defined by \cite{Porrati:2001db}%
\footnote{Some signs have been changed to conform to our conventions.}
\begin{equation}
\Delta_0=-\fft1{4\pi^2L^2}\left(\alpha\fft1{Z^2-1}-\beta\fft{Z}{Z^2-1}
\right)
\end{equation}
we find $\alpha_+=(\alpha+\beta)$ and $\alpha_-=-(\alpha-\beta)$.

For the fermion propagator, we consider the Dirac equation in
homogeneous coordinates.  Start by defining the Dirac operator
$K=\Gamma^{MN}X_M\partial_N$ where $\{\Gamma^M,\Gamma^N\}=2\eta^{MN}$.
By squaring this operator, it is easy to show that
$K(K-3)=\hat N(\hat N+3)-X^2\partial^2$.  On the other hand, by squaring
the $SO(2,3)$ generators acting on a spin-$\ha$ state,
$L_{MN}=i(X_M\partial_N-X_n\partial_M)+\fft{i}2\Gamma_{MN}$, we may
show that $Q=\hat N(\hat N+3)-X^2\partial^2+\fft52-K=K(K-4)+\fft52$.
When acting on $\Psi(X)$, this must reproduce the Casimir
$Q=E_0(E_0-3)+\fft34$.  Equating these expressions, we find the
factorized relation $(K-\ha)(K-\fft72)=E_0(E_0-3)$, so that either
$E_0=K-\ha$ or $E_0=\fft72-K$.  This gives two possible Dirac equations
\begin{equation}
[K-(E_0+\half)]\Psi(X)=0 \qquad\hbox{or}\qquad [K+(E_0-\ft72)]\Psi(X)=0
\label{eq:dire}
\end{equation}
For the massless case ($E_0=\fft32$), both equations degenerate to
$(K-2)\Psi(X)=0$.

Next, we note the factorization $(K-\lambda)(K+\lambda-3)=\hat
N(\hat N+3)-X^2\partial^2-\lambda(\lambda-3)$, which holds for arbitrary
$\lambda$.  Since the right hand side is simply the scalar Klein-Gordon
operator, (\ref{eq:kge}), this provides the AdS equivalent of the
relation $(/\kern-6pt\partial-m)(/\kern-6pt\partial+m)=\square-m^2$.
Denoting either $\lambda$ or $3-\lambda$ by $E_0^{(0)}$ (indicating the
canonical value of $E_0$ in the scalar equation), this may be rewritten in
the suggestive manner \cite{Janssen:dr}
\begin{eqnarray}
[K-(E_0+\half)][K+(E_0-\ft52)]&=&\hat N(\hat
N+3)-X^2\partial^2-E_0^{(0)}(E_0^{(0)}-3),\qquad
E_0^{(0)}=E_0+\half\nonumber\\ \relax
[K+(E_0-\ft72)][K-(E_0-\half)]&=&\hat N(\hat
N+3)-X^2\partial^2-E_0^{(0)}(E_0^{(0)}-3),\qquad
E_0^{(0)}=E_0-\half\nonumber\\
\end{eqnarray}
so that solutions to the Dirac equation, (\ref{eq:dire}), are easily
obtained from solutions to the scalar equation, (\ref{eq:kge}), by
taking
\begin{eqnarray}
\Psi(X)&=&[K+(E_0-\ft52)]\Psi_0\,\phi(X;E_0^{(0)}=E_0+\half)\nonumber\\
\noalign{\noindent or}
\Psi(X)&=&[K-(E_0-\half)]\Psi_0\,\phi(X;E_0^{(0)}=E_0-\half)
\label{eq:drel}
\end{eqnarray}
with $\Psi_0$ a constant spinor.  This result allows us to immediately
determine the fermion propagator in terms of the scalar one in
much the same way as one would compute $1/\dslash = \dslash/\,\square$
in the flat limit.

For $E_0=\fft32$ (corresponding to $E_0^{(0)}=1$ or $2$), we use the
form of the scalar propagator, (\ref{eq:aatscal}), and the relation
(\ref{eq:drel}), to obtain
\begin{equation}
\Delta_{1/2}=\fft1{8\pi^2L^4}\fft{\Gamma^M(X_M-Y_M)}{(Z+1)^2}=\fft1{2\pi^2}
\fft{\Gamma^M(X_M-Y_M)}{|X-Y|^4}
\end{equation}
This is the massless fermion propagator corresponding to transparent
boundary conditions.  Similarly to Eq.~(\ref{eq:aagscal}), general boundary
conditions may be imposed by introducing parameters $\alpha_+,\alpha_-$
and taking
\begin{equation}
\Delta_{1/2}^{(\alpha)}=\fft1{8\pi^2L^4}\left(\alpha_+\fft{\Gamma^M(X_M-Y_M)}
{(Z+1)^2}
+\alpha_-\fft{\Gamma^M(X_M+Y_M)}{(Z-1)^2}\right)
\label{eq:aa12prop}
\end{equation}

Turning next to the vector propagator, we use the results of
Ref.~\cite{Allen:wd}, converted to homogeneous coordinates.  The vector
propagator is the first case where we have to worry about bi-tensor
structures as well as gauge fixing.  However, fortunately, for correlators
of the stress tensor, we only need
the expression for the gauge invariant two-point function $\langle
F_{MN}(X)F_{PQ}(Y)\rangle$.  Based on symmetry, this expression can be
written as
\begin{equation}
\langle F_{MN}(X)F^{PQ}(Y)\rangle=\sigma(Z)\hat G_{[M}{}^{[P}\hat G_{N]}{}^{Q]}
+\tau(Z)N_{[M}\hat G_{N]}{}^{[Q}N^{P]}
\end{equation}
where $\sigma(Z)$ and $\tau(Z)$ may be determined as in
Ref.~\cite{Allen:wd}.  Taking into account mixed boundary conditions as
well as normalization of the short distance behavior, we find
\begin{eqnarray}
\langle F_{MN}(X)F^{PQ}(Y)\rangle^{(\alpha)}&=&\fft1{2\pi^2L^4}\Biggl[
\fft{\alpha_+}{(Z+1)^2}
(\hat G_{[M}{}^{[P}\hat G_{N]}{}^{Q]}-2(Z-1)N_{[M}\hat G_{N]}{}^{[Q}N^{P]})
\nonumber\\
&&\qquad\quad
+\fft{\alpha_-}{(Z-1)^2}
(\hat G_{[M}{}^{[P}\hat G_{N]}{}^{Q]}-2(Z+1)N_{[M}\hat G_{N]}{}^{[Q}N^{P]})
\Biggr]
\label{eq:aa1prop}
\end{eqnarray}
These mixed boundary condition propagators, (\ref{eq:aagscal}),
(\ref{eq:aa12prop}) and (\ref{eq:aa1prop}), are the ones used in the
one-loop computation.

%%%%%%%%%%%%%%%%%%%%%%%%%%%%%%%%%%%%%%%%%%%%%%%%%%%%%%%%%%%%%%%%%%%%%%%%%%%%%%

\section{A transverse-traceless bi-tensor basis}

In this appendix, we present a convenient basis into which any
transverse-traceless bi-local tensor may be decomposed.
Since any traceless tensor, $\cal T$,
may be decomposed in terms of the three $T$ tensors defined in
(\ref{eq:ttens}), we start by writing ${\cal T}=a_1(3Z^2+1)T_1+a_2T_2+a_3T_3$.
The factor $(3Z^2+1)$ is introduced for convenience.  We now impose
transversality on $\cal T$.  In particular, taking the
divergence of $\cal T$ on the first index gives
\begin{eqnarray}
\nabla^M{\cal T}_{MNQP}&=&\fft{\sqrt{Z^2-1}}{L}\Bigl[((3Z^2+1)a_1'+6Za_1)
N\cdot T_1+a_2'N \cdot T_2+a_3'N\cdot T_3\Bigr]\nonumber\\
&&+a_1(3Z^2+1)\nabla^MT_1+a_2\nabla^MT_2+a_3\nabla^MT_3
\label{eq:dpi}
\end{eqnarray}
We compute
\begin{equation}
N\cdot T_1=\fft1{3Z^2+1}A,\qquad
N\cdot T_2=0,\qquad
N\cdot T_3=\fft1{2Z}B
\end{equation}
and
\begin{eqnarray}
\sqrt{Z^2-1}\nabla\cdot
T_1&=&\fft{2Z(3Z^2+5)}{(3Z^2+1)^2}A
-\fft4{3(3Z^2+1)}
B\nonumber\\
\sqrt{Z^2-1}\nabla\cdot
T_2&=&-\fft53B\nonumber\\
\sqrt{Z^2-1}\nabla\cdot
T_3&=&\fft{3Z^2+1}{2Z^2}
B
-\fft1Z A
\end{eqnarray}
where the tensors $A$ and $B$ are given by
\begin{equation}
A_{NPQ}=(4N_NN_PN_Q-N_NG_{PQ}),
\qquad
B_{NPQ}=((\hat G_{NP}N_Q+\hat G_{NQ}N_P)-2ZN_NN_PN_Q)
\end{equation}
Thus the vanishing of the divergence in (\ref{eq:dpi}) leads to two
conditions on the three functions
\begin{eqnarray}
(Z^2-1)Za_1'&=&-4Z^2a_1+a_3\nonumber\\
(Z^2-1)Za_3'&=&\ft83Z^2a_1+\ft{10}3Z^2a_2-(3Z^2+1)a_3
\end{eqnarray}
These equations may be solved to give $a_2$ and $a_3$ in terms of $a_1$
and its derivatives.  As a result, any transverse traceless bi-tensor must
take the form
\begin{eqnarray}
{\cal T}&=&a(3Z^2+1)T_1+[\ft3{10}(Z^2-1)^2a''+3(Z^2-1)Za'+2(3Z^2-1)a]T_2
\nonumber\\
&&\quad+[(Z^2-1)Za'+4Z^2a]T_3
\label{eq:ttcond}
\end{eqnarray}
and is fully specified by the function $a(Z)$.

By choosing a complete set of functions $a(Z)$, we may obtain a basis of
transverse traceless bi-tensors.  A convenient choice is to take
$a_{(n)}=1/Z^n$, whereupon the resulting expression of (\ref{eq:ttcond})
may be denoted ${\cal T}_{(n)}$.  The first few basis bi-tensors with
sufficiently fast large distance falloff are shown in Table~\ref{tbl:1}.
Note the absence of leading order $1/Z^2$ and $1/Z^3$ behavior in the
$a_2$ and $a_3$ coefficients of ${\cal T}_{(4)}$ and ${\cal T}_{(5)}$,
respectively.

%%%%%%%%%%%%%%%%%%%%%%%%%%%%%%%%%%%%%%%%%%%%%%%%%%%%%%%%%%%%%%%%%%%%%%%%%%%%%%
%% Tables

\begin{table}
\begin{tabular}{llll}
&$a_1$&$a_2$&$a_3$\\
\hline
${\cal T}_{(4)}$&$Z^{-4}$&$-2Z^{-4}+6Z^{-6}$&$4Z^{-4}$\\
${\cal T}_{(5)}$&$Z^{-5}$&$-5Z^{-5}+9Z^{-7}$&$-Z^{-3}+5Z^{-5}$\\
${\cal T}_{(6)}$&$Z^{-6}$&$\fft35Z^{-4}-\fft{46}5Z^{-6}
+\fft{63}5Z^{-8}$&$-2Z^{-4}+6Z^{-6}$\\
${\cal T}_{(7)}$&$Z^{-7}$&$\fft95Z^{-5}-\fft{73}5Z^{-7}
+\fft{84}5Z^{-9}$&$-3Z^{-5}+7Z^{-7}$\\
\end{tabular}
\caption{First few elements of the transverse traceless bi-tensor basis
${\cal T}_{(n)}$.  The coefficients $a_1$, $a_2$ and $a_3$ correspond to
the decomposition ${\cal T}=a_1(3Z^2+1)T_1+a_2T_2+a_3T_3$.}
\label{tbl:1}
\end{table}

%%%%%%%%%%%%%%%%%%%%%%%%%%%%%%%%%%%%%%%%%%%%%%%%%%%%%%%%%%%%%%%%%%%%%%%%%%%%%%
%% References


\begin{references}

\bibitem{vdv}
H.~van Dam and M.~Veltman,
\rtitle{Massive and massless Yang-Mills and gravitational fields}
Nucl.\ Phys.\ {\bf B22}, 397 (1970).
%%CITATION = NUPHA,B22,397;%%

\bibitem{zak}
V.I. Zakharov, JETP Lett. {\bf 12}, 312 (1970).
%%We did not have this you might want to check it.%%
%%CITATION = JTPLA,12,312;%%

\bibitem{Porrativvz}
M.~Porrati,
\rtitle{No van Dam-Veltman-Zakharov discontinuity in AdS space}
Phys.\ Lett.\ B {\bf 498}, 92 (2001) [hep-th/0011152].
%%CITATION = HEP-TH 0011152;%%

\bibitem{kogan}
I.I.~Kogan, S.~Mouslopoulos and A.~Papazoglou,
Phys.\ Lett.\ B {\bf 503}, 173 (2001) [hep-th/0011138].
%%CITATION = HEP-TH 0011138;%%

\bibitem{duff:dls}
M.J.~Duff, J.T.~Liu and H.~Sati,
\rtitle{Quantum $M^{2} \rightarrow 2\Lambda/3$ discontinuity for massive
gravity with a $\Lambda$ term}
Phys. Lett. {\bf B516} (2001) 156 [hep-th/0105008].
%%CITATION = HEP-TH 0105008;%%

\bibitem{duff:ddls}
F.A.~Dilkes, M.J.~Duff, J.T.~Liu and H.~Sati,
\rtitle{Quantum discontinuity between zero and infinitesimal graviton mass}
Phys.  Rev.  Lett.  {\bf 87} (2001) 041301 [hep-th/0102093].
%%CITATION = HEP-TH 0102093;%%

\bibitem{Duffdyna}
M.J. Duff,
\rtitle{Dynamical breaking of general covariance and massive spin-two mesons}
Phys. Rev. {\bf D12}, 3969 (1975).
%%CITATION = PHRVA,D12,3969;%%

\bibitem{Porrati:2001db}
M.~Porrati,
\rtitle{Higgs phenomenon for 4-D gravity in anti de Sitter space}
JHEP {\bf 0204}, 058 (2002) [arXiv:hep-th/0112166].
%%CITATION = HEP-TH 0112166;%%

\bibitem{karch}
A.~Karch and L.~Randall,
\rtitle{Locally localized gravity}
JHEP {\bf 0105}, 008 (2001) [arXiv:hep-th/0011156].
%%CITATION = HEP-TH 0011156;%%

\bibitem{Porrati:2001gx}
M.~Porrati,
\rtitle{Mass and gauge invariance. IV: Holography for the Karch-Randall model}
Phys.\ Rev.\ D {\bf 65}, 044015 (2002) [arXiv:hep-th/0109017].
%%CITATION = HEP-TH 0109017;%%

\bibitem{Bousso:2001cf}
R.~Bousso and L.~Randall,
JHEP {\bf 0204}, 057 (2002) [arXiv:hep-th/0112080].
%%CITATION = HEP-TH 0112080;%%

\bibitem{Duffliu}
M.J.~Duff and J.T.~Liu,
Phys.\ Rev.\ Lett.\ {\bf 85}, 2052 (2000)
[Class.\ Quant.\ Grav.\ {\bf 18}, 3207 (2001)] [arXiv:hep-th/0003237].
%%CITATION = HEP-TH 0003237;%%

\bibitem{Maldacena}
J. Maldacena,
Adv. Theor. Math. Phys. {\bf 2}, 231 (1998)
[Int.\ J.\ Theor.\ Phys.\  {\bf 38}, 1113 (1999)] [arXiv:hep-th/9711200].
%%CITATION = HEP-TH 9711200;%%

\bibitem{Wittenads}
E. Witten,
Adv. Theor. Math. Phys. {\bf 2}, 253 (1998) [arXiv:hep-th/9802150].
%%CITATION = HEP-TH 9802150;%%

\bibitem{Gubserklebanovpolyakov}
S.S. Gubser, I.R. Klebanov and A.M. Polyakov,
Phys. Lett. B {\bf 428}, 105 (1998) [arXiv:hep-th/9802109].
%%CITATION = HEP-TH 9802109;%%

\bibitem{Randall}
L. Randall and R. Sundrum,
Phys. Rev. Lett. {\bf 83}, 4690 (1999) [arXiv:hep-th/9906064].
%%CITATION = HEP-TH 9906064;%%

\bibitem{Duff2}
M. J. Duff,
\rtitle{Quantum corrections to the Schwarzschild solution}
Phys. Rev. {\bf D9}, 1837 (1974).
%%CITATION = PHRVA,D9,1837;%%

\bibitem{Duffweyl}
M. J. Duff,
\rtitle{Twenty years of the Weyl anomaly}
Class. Quant. Grav. {\bf 11}, 1387 (1994) [arXiv:hep-th/9308075].
%%CITATION = HEP-TH 9308075;%%

\bibitem{DeWolfe:2001pq}
O.~DeWolfe, D.Z.~Freedman and H.~Ooguri,
Phys.\ Rev.\ D {\bf 66}, 025009 (2002)
[arXiv:hep-th/0111135].
%%CITATION = HEP-TH 0111135;%%

\bibitem{Erdmenger:2002ex}
J.~Erdmenger, Z.~Guralnik and I.~Kirsch,
arXiv:hep-th/0203020.
%%CITATION = HEP-TH 0203020;%%

\bibitem{Fronsdal:1978vb}
C.~Fronsdal,
%% \rtitle{Singletons and massless, integral-spin fields on de Sitter space}
Phys.\ Rev.\ D {\bf 20}, 848 (1979).
%%CITATION = PHRVA,D20,848;%%

\bibitem{Allen:wd}
B.~Allen and T.~Jacobson,
\rtitle{Vector Two Point Functions In Maximally Symmetric Spaces}
Commun.\ Math.\ Phys.\  {\bf 103}, 669 (1986).
%%CITATION = CMPHA,103,669;%%

\bibitem{Allen:1986tt}
B.~Allen and M.~Turyn,
\rtitle{An Evaluation Of The Graviton Propagator In De Sitter Space},
Nucl.\ Phys.\ B {\bf 292}, 813 (1987).
%%CITATION = NUPHA,B292,813;%%

\bibitem{Turyn:1988af}
M.~Turyn,
\rtitle{The Graviton Propagator In Maximally Symmetric Spaces}
J.\ Math.\ Phys.\  {\bf 31}, 669 (1990).
%%CITATION = JMAPA,31,669;%%

\bibitem{D'Hoker:1999jc}
E.~D'Hoker, D.~Z.~Freedman, S.~D.~Mathur, A.~Matusis and L.~Rastelli,
\rtitle{Graviton and gauge boson propagators in AdS$_{d+1}$}
Nucl.\ Phys.\ B {\bf 562}, 330 (1999) [arXiv:hep-th/9902042].
%%CITATION = HEP-TH 9902042;%%

\bibitem{Avis:1977yn}
S.J.~Avis, C.J.~Isham and D.~Storey,
%% \rtitle{Quantum Field Theory In Anti-De Sitter Space-Time}
Phys.\ Rev.\ D {\bf 18}, 3565 (1978).
%%CITATION = PHRVA,D18,3565;%%

\bibitem{Breitenlohner:jf}
P.~Breitenlohner and D.~Z.~Freedman,
\rtitle{Stability In Gauged Extended Supergravity}
Annals Phys.\  {\bf 144}, 249 (1982).
%%CITATION = APNYA,144,249;%%

\bibitem{Janssen:dr}
H.~Janssen and C.~Dullemond,
\rtitle{Spinor Propagators In Anti-De Sitter Space-Time}
J.\ Math.\ Phys.\  {\bf 27}, 2786 (1986).
%%CITATION = JMAPA,27,2786;%%

\end{references}
\end{document}